\documentclass{aa}  

\usepackage{graphicx}
\usepackage{blindtext}
\usepackage{txfonts}
\usepackage[colorlinks=true,linkcolor=blue,citecolor=blue,urlcolor=blue]{hyperref}
\usepackage{comment}
\usepackage{dcolumn}
\usepackage{arydshln} 
\usepackage{natbib}  

\usepackage{gensymb}
\usepackage{subcaption}
\usepackage{linenoaa}
\usepackage{xcolor}

\begin{document}

   \title{Magnetic connectivity from the Sun to the Earth with MHD models}

   \subtitle{II. Impact of input magnetic maps}

   \author{B. Perri  \inst{1}
          \and
          S. Kennis   \inst{2}
          \and 
          B. Schmieder \inst{2,4,5}
          \and 
          S. Poedts \inst{2,3}
         }

   \institute{AIM/DAp - CEA Paris-Saclay, Université Paris-Saclay, Université Paris-Cité, Gif-sur-Yvette, France
              \email{barbara.perri@cea.fr}
         \and
         Centre for Mathematical Plasma-Astrophysics, KU Leuven, Celestijnenlaan 200B, 3001 Leuven, Belgium       
         \and 
         Institute of Physics, University of Maria Curie-Skłodowska, Pl.\ M.\ Curie-Skłodowska 5, 20-031 Lublin, Poland
         \and
         Observatoire de Paris, LIRA, UMR8254 (CNRS), F-92195 Meudon Principal Cedex, France
         \and
         LUNEX EMMESI COSPAR-PEX Eurospacehub, Kapteyn straat 1, Noordwijk 2201 BB, Netherlands}

   \date{} 
 
  \abstract  
   {Magnetic connectivity is a method used to understand the connection between the Sun and the satellites orbiting around it. It is especially useful for Earth, as it helps us anticipate the propagation and geoeffectiveness of space weather events. Connectivity estimation needs to be fast (within 24 hours) and thus usually relies on semi-empirical models. In the first paper of this study, we established that optimised magnetohydrodynamics (MHD) models can also be used to derive magnetic connectivity, yielding good results, especially for complex configurations.}
   {However, the modelling is not the only factor that can impact the connectivity estimation. Models' results, both semi-empirical and MHD, depend heavily on the input data used to drive the simulation. The impact of the chosen input magnetic map has been studied for coronal and heliospheric models independently, but rarely together, and with a specific focus on magnetic connectivity. This knowledge is crucial to reducing the uncertainty associated with space weather forecasts.}
   {In order to quantify the impact of input magnetic maps on our connectivity estimation, we used twenty different maps by four providers, with various processing treatments. We selected the period of July 2019, which corresponds to Carrington rotation (CR) number 2219, because it had already been studied with COCONUT in a previous paper. It corresponds to a period of minimum activity, meaning minimal interference from transient structures, but with a high-speed stream and several polarity switches, making it interesting and challenging to model. We have compared our results with in situ data from ACE and developed several visualisation techniques to validate connectivity not only for a specific date but also for a full CR.}
   {The choice of the input magnetic map drastically affects the connectivity estimates, with global scores going from 28\% global agreement to 63\% in the most extreme case. The HMI and GONG maps consistently gave the best connectivity estimations, with more than 50\% correct results of the full CR. For this specific period, the GONG-ADAPT maps yielded the least reliable results, with a score below 35\%. This is not surprising, as we have shown before that COCONUT required parameter adjustments for these maps. However, WSO maps which showed poor coronal agreement actually provided good connectivity estimates, showing that the Heliospheric Current Sheet (HCS) location is in the end the most crucial parameter for connectivity (compared to open field areas locations or streamer shape).}
   {We have confirmed with this study that minimum of activity cases are actually challenging for magnetic connectivity, due to the proximity of the HCS with Earth's position. Consistent data over a full Carrington rotation are to be favored compared to daily updates (which proved more unstable in terms of connectivity), or even high-resolution.}


   \maketitle
   \nolinenumbers

\section{Introduction}
Connectivity refers to the capacity to connect in situ observations measured around a spacecraft with the solar region responsible for the physical phenomena being measured. For instance, in the case of transient events, energetic particles measured at Earth can be traced back to the solar flare that accelerated them \citep{Roelof1973}. Similarly, for continuous connections between the Sun and the Earth, a fast solar wind stream at Earth can be traced back to a coronal hole in the solar atmosphere \citep{Krieger1973}. 
Connectivity is critical for space weather and our general understanding of the Sun-Earth chain, as it allows us to put different kinds of observations into perspective and to retrieve a global view of the physical events. 
However, it is currently very difficult to achieve this. Indeed, there are no complete 3D observations of the Sun-Earth line (although complementary viewpoints such as PUNCH at L1 and Vigil at L5 will help in the future). For transients, we can work around this issue by using radio observations: radio bursts can trace the propagation of particles accelerated by flares and Coronal Mass Ejections (CMEs) in the interplanetary medium by measuring the Langmuir waves generated along the path \citep{Chen2013, Morosan2021}, although it has been shown that type-III radio bursts can also appear without connection to solar flares, so it is not a one-to-one correspondance \citep{Harra2021}. For the solar wind, we can analyze the composition of the studied stream: it has been shown that the solar wind composition is frozen above the corona and depends on the magnetic configuration in the chromosphere, hence differentiating open and closed magnetic regions \citep{Landi2012, Varesano2024, Wallace2025}. Unfortunately, none of these methods is systematic at the moment, and hence cannot be used on a large number of cases or in an operational space weather setting.  

This is why we turn to modelling to estimate the solar connectivity. With this method, we can compute two physical quantities related to the Sun: the solar wind speed and the radial component of the interplanetary magnetic field (IMF). Indeed, solar wind streams tend to have a consistent acceleration along their trajectories, while we know that the IMF is organized at large scales by the Parker spiral structure \citep{Parker1958}. If the magnetohydrodynamics (MHD) approximations are valid, plasma and magnetic structures act as one due to the frozen-in field approximation. The magnetic field structure is especially important since most transients are sensitive to its topology \citep{Heber2006}. 
Hence, the most commonly used method to determine the solar connectivity is the two-step ballistic mapping \citep{Peleikis2017}: close to the Sun, the magnetic field is extrapolated based on the potential field source surface (PFSS) approximation \citep{Schatten1969} (which is called magnetic mapping), and further away in the interplanetary medium, it is connected to the Parker spiral (which is called ballistic mapping). An operational implementation of this method is available through the Magnetic Connectivity Tool\footnote{\url{https://connect-tool.irap.omp.eu/}} \citep{Rouillard2020}. This method has the advantage of being fast and robust, but it has many free parameters that can affect and even change the final result \citep{Badman2020, Badman2023}. In particular, the source surface of the PFSS or the angle of the Parker spiral related to the chosen wind speed are known to have a strong impact on the final connectivity estimate \citep{koukras2022, DaSilva2023, Dakeyo2024}. Also, the ballistic mapping is a highly idealized view of the IMF, which, in reality, can be distorted by turbulence, waves, shears, shocks, or magnetic reconnection \citep{Owens2013}. 
This is why we can go beyond this method and directly use MHD models to compute connectivity: by tracing magnetic field lines within one or several models, we can take a more physically grounded approach (although more computationally expensive, and still limited by the choices of physics modules included and modelling simplifications that can be made). Our previous paper \citep{Kennis2024}, hereafter referenced as Paper~Ia, showed that the MHD chain COCONUT \citep{Perri2022} + EUHFORIA \citep{Pomoell2018} (or COCORIA for the full chain, see \cite{Linan2025}) yielded good results for the connectivity in various previously studied cases \citep{Reiss2021, koukras2022}, outperforming even the two-step ballistic method for the most complex cases\footnote{A direct comparison with the Magnetic Connectivity Tool could not be made for most of the cases as they happened before 2020, but we developed our own two-step ballistic algorithm by coupling a PFSS extrapolation to a Parker spiral model, which yields very similar results.}. In particular, we estimated the uncertainties associated with both methods, computing spatial and temporal uncertainties for the COCORIA chain.  

However, there is one type of uncertainty we want to examine in more detail in this study: the uncertainty associated with the choice of the input magnetogram. As discussed in another previous paper \citep{Perri2023}, hereafter Paper~Ib, there are many magnetic maps available for each date, and they will all yield different results for solar wind and magnetic field estimates, even during periods of minimal activity. This is due to the various parameters and approximations used to reconstruct these synoptic maps: resolution, the combination of non-synchronic data and polar extrapolations, and the choice of coronal model are the most critical factors \citep{riley2014}. As shown in Paper~Ib, the choice of the input map has a huge impact on the coronal structure, as it can change the disposition of open magnetic field areas, the position of the streamers, and consequently the tilt of the heliospheric current sheet (HCS). All of these parameters will affect connectivity through the IMF structure. 
By connecting the COCONUT simulations from Paper~Ib to EUHFORIA, we can more precisely quantify the impact of the input magnetic map on Earth's connectivity at 1 astronomical unit (AU).

The structure of this paper is as follows. In section~\ref{sec:recap_chain}, we describe the modelling chain starting from the magnetic maps, and sum up the main properties of the MHD models we use. In section~\ref{sec:connectivity}, we discuss the connectivity computation and present the new analysis developed for this paper. In Section~\ref{sec:results}, we present the validation of each solution for each magnetic map, and discuss the differences and their origins in Section \ref{sec:discussion}. At last, the conclusions are presented in Section~\ref{sec:conclusion}.

\section{Description of the modelling chain}
\label{sec:recap_chain}

We recall here how the full COCORIA modelling chain works. We start with the description of the input magnetic maps in Section~\ref{sec:maps}. Then we recap the main properties of the MHD models COCONUT and EUHFORIA in Section~\ref{sec:mhd_models}. For more details about the models, the reader can refer to Paper~Ia.  

\subsection{Input magnetic maps}
\label{sec:maps}

As explained in the introduction, we use the same input magnetic maps as in Paper~Ib. Hence, all the technical details are in \cite{Perri2023}, and we will just recap the main differences between the maps here. 

We selected 20 synoptic magnetic maps for the date of July 2nd, 2019, from 4 different providers: the Helioseismic and Magnetic Imager (HMI, an instrument on the Solar Dynamics Observatory), the Global Oscillation Network Group (GONG), the 'Air Force Data Assimilative Photospheric Flux transport' (ADAPT) and the Wilcox Solar Observatory (WSO). This date was originally selected because it corresponded to a total solar eclipse, which made the comparison with coronal observations easier. It corresponds to Carrington rotation number 2219, which runs from June 29, 2019, to July 26, 2019. 

We start with the highest-resolution maps from HMI. 
The resolution of these maps is 3600 in longitude and 1440 in sine latitude, which is possible because the observations are space-based. 
It focuses on the Fe line at a wavelength of 617.3~nm. 
They are based on 720s line-of-sight magnetograms converted to remapped radial-field magnetograms. The closest pixels to the central meridian are extracted for each longitude and put together using a Gaussian function. 
The polar fields are interpolated both temporally and spatially \citep{Sun2011}.
HMI standard maps are produced in the Carrington frame, once per Carrington rotation. 
In contrast, the HMI daily maps are in a synchronic frame and are updated daily. This means that for each moment, the map between a longitude of 0$^{\circ}$ and 120$^{\circ}$ is replaced by the magnetogram observation of that moment between 10 and 14 UT.

Next, we have the Global Oscillation Network Group (GONG). 
The resolution of these maps is 360 in longitude and 180 in sine latitude. 
The spectral line used to observe the Sun is Ni at a wavelength of 676.8~nm. 
There are 5 different synoptic maps available. Two of them are in the Carrington frame: the \textit{mrmqs} and \textit{mrnqs} maps. They are built using fully calibrated one-minute full-disk remapped photospheric magnetograms from GONG's six ground-based sites, merged with weighting factors in the form of a cosine to the power of 4 of the longitude.\footnote{\url{https://gong.nso.edu/data/dmac_magmap/}} 
The \textit{mrnqs} is a zero-point-corrected product; these maps include corrections at the poles to improve the estimate of the global magnetic flux.
The 3 others are synchronic-frame maps. The \textit{mrbqj} product (Janus) is similar to the HMI synchronic frame maps: the left 60 degrees of longitude, between -60 and 60 degrees in latitude, are updated. The \textit{mrbqs} and \textit{mrzqs} products are in the synchronic frame but are not updated. The \textit{mrzqs} is also a zero-point-corrected product. That makes the \textit{mrbqs} product the standard one. 
For all maps, the polar fields are interpolated using a cubic-polynomial fit. 

The ADAPT synoptic maps are based on other synoptic maps, such as GONG, HMI, vsm or kpvt. However, when this study was conducted, only GONG-ADAPT maps were released for the time period we considered, so we will focus only on this product in this paper. GONG-ADAPT maps are thus based on GONG data, so they have the same resolution (360 in longitude, 180 in latitude) and are observed at the same spectral line. However, the combination of magnetograms differs here. These synoptic maps are produced using the surface flux transport model and data assimilation, which means that some of the information comes from simulations and some from observations \citep{Hickmann2015}. The main consequence is that they have a larger magnetic flux at the poles than other maps \citep{Huichao2021}. Because of this additional modelling phase, ADAPT maps come with 12 realisations for each date and hour to test the effects of certain parameters within the model.

The last type of map is from WSO (Wilcox Solar Observatory). This synoptic map is constructed using data from different magnetic maps obtained by observing the Fe I line. This particular type of synoptic map is derived from ground-based data, which results in a lower resolution (73 in longitude, 30 in latitude or sine latitude, and only from -70 to 70 degrees). A linear extrapolation is suggested to fill the polar fields.  

For more details, the reader can also refer to the reviews of \cite{riley2014} and \cite{virtanen2017}.

\subsection{MHD models}
\label{sec:mhd_models}

Instead of using a single model to compute the Sun-Earth chain, we use a coronal model from the solar surface to 0.1~AU, followed by a heliospheric model from 0.1~AU to 2 AU for computational efficiency. 
As explained in the introduction, this study is based on the same modelling chain as in Paper~Ia. Hence, all technical details can be found in \cite{Kennis2024}; we recall only the most important elements here.  

The coronal model is COCONUT (COolfluid COroNa UnsTructured, see \cite{Perri2022}). It is a 3D MHD data-driven model, based on the COOLFluid framework \citep{Lani2013}. It solves the ideal MHD equations written in their conservative form in Cartesian coordinates and dimensionless form. It uses a time-implicit backwards-Euler scheme in finite-volume methods to achieve operational running times \citep{Perri2022} and runs on an unstructured grid to avoid polar singularities \citep{Brchnelova2022}. The default mesh used a 6th-level subdivision of the geodesic polyhedron, yielding 20,480 surface elements and a grid of 3.9M elements. To ensure the divergence constraint $\nabla \cdot \boldsymbol{B} = 0$, we use the Artificial Compressibility Analogy \citep{chorin1997}.
The model neglected certain terms to speed up the calculations, such as the Hall term in the induction equation. Also, the Coriolis force or centrifugal forces are neglected in the momentum equation. 
The input map is preprocessed by the code before the simulation: it is projected onto spherical harmonics and reconstructed using the first 30 modes. 
This version of the COCONUT model uses a polytropic approximation (with a polytropic index of 1.05). The model was run with the same parameters described in \cite{Perri2023}, which means $\rho_\odot = 1.67 \times 10^{-16}\ \mathrm{g/cm^3}$ and $T_\odot = 1.9 \times 10^6\ \mathrm{K}$ for fixed-value Dirichlet conditions of density and pressure. The pressure at the inner boundary follows from the solar surface temperature by application of the ideal gas law: $P_\odot = 4.15 \times 10^{-2} \, \mathrm{dyn/cm^2}$. All other magnetic field quantities satisfy zero-gradient conditions. As for the velocity, it is aligned with the magnetic field to limit surface currents, as explained in \citet{Brchnelova2022b}. 
The frame used in COCONUT is the same as the input map (either Carrington or synchronic). 
The interface with the heliospheric domain is done by providing the solar wind radial speed, temperature, density and radial magnetic field at 0.1~AU.

The heliospheric model is that of EUHFORIA (EUropean Heliospheric FORecasting Information Asset; see \cite{Pomoell2018}). It is a 3D model in which the ideal MHD equations, including gravity, are solved using a finite-volume method combined with a constrained-transport approach. We use an approximate Riemann solver with standard piecewise-linear reconstruction \citep{Kissmann2012, Pomoell2012} to obtain a robust, second-order-accurate scheme. We choose to omit the Coriolis and centrifugal terms.
A value of 1.5 is selected for the polytropic index as in \cite{Odstrcil2004}.
The default setup has a 2$\degree$ angular resolution.
At the outer radial boundary, we use open boundary conditions implemented via a simple extrapolation, whereas at the latitudinal boundaries, we use symmetric reflection boundary conditions.
The frame is the Heliocentric Earth Equatorial coordinate system (HEEQ-coordinates).  

The full modelling chain comprising COCONUT as a coronal model and EUHFORIA as a heliospheric model is hereafter referred to as COCORIA. It can also be used with CME injection \citep{Linan2025}, but we will only use the background solar wind in this study. 

\section{Computation of the magnetic connectivity}
\label{sec:connectivity}

We recall here how we compute the magnetic connectivity using the COCORIA modelling chain. 
In Section \ref{sec:recap_connectivity}, we summarize the method implemented in Paper~Ia. 
In Section~\ref{sec:connectivity_carrington}, we explain the new developments used in this study to visualize the connectivity for a full Carrington rotation. 
Finally, in Section \ref{sec:ace}, we describe the expected output data at Earth based on observations from the ACE spacecraft.

\subsection{Connectivity snapshot}
\label{sec:recap_connectivity}

\begin{figure*}[!h]
    \centering
    \includegraphics[width=\hsize]{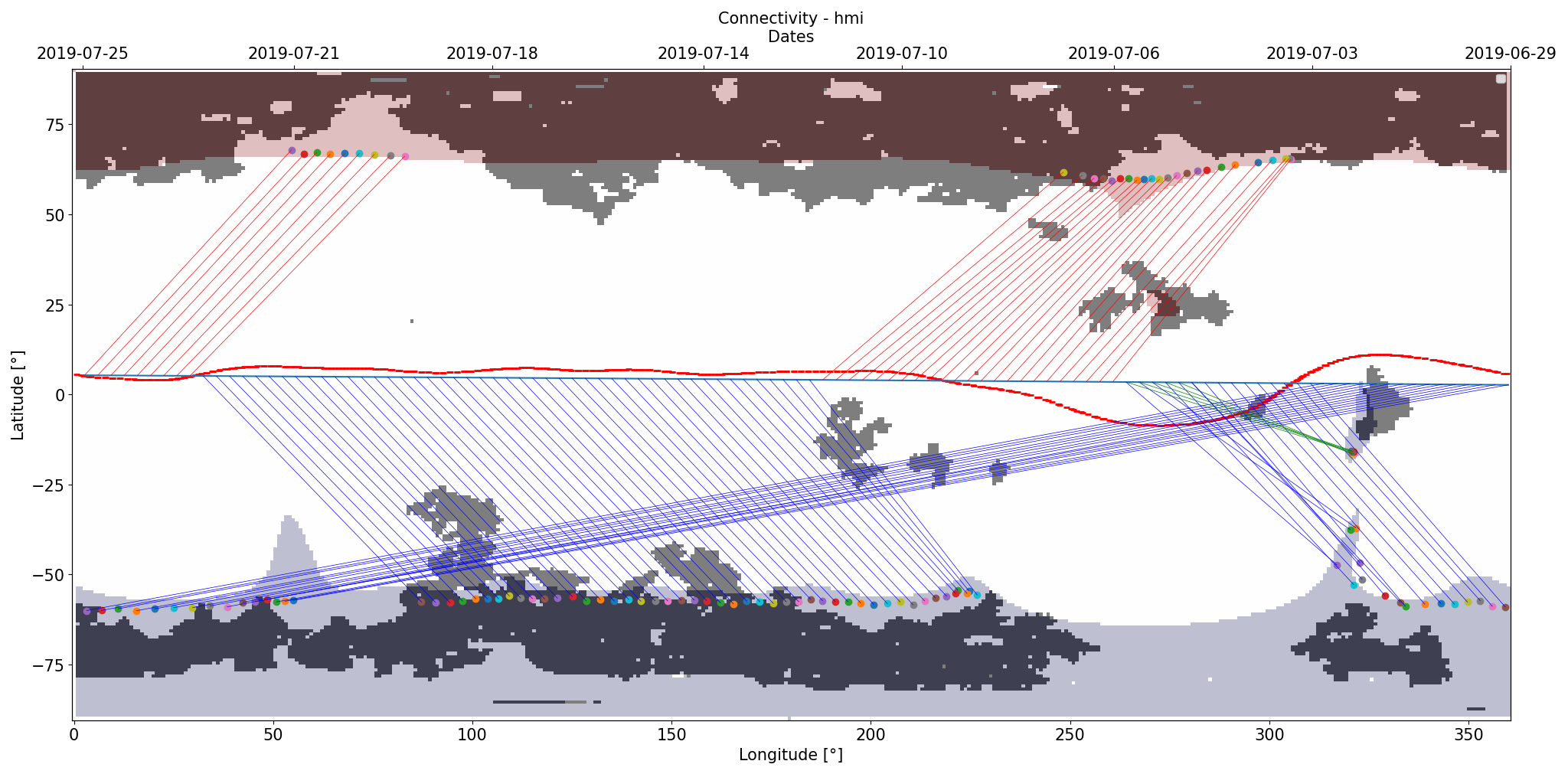}
    \caption{Visualization of the magnetic connectivity between the Sun and the Earth for the full Carrington rotation number 2219 for the HMI synoptic map. The Earth's trajectory over the period is shown as a horizontal blue line. For each position, connectivity to a positive polarity open field area is shown with red lines, while connectivity to a negative polarity open field area is shown with blue lines, and connectivity to an equatorial open field area is shown with green lines. The open field areas of the corresponding COCORIA simulation are shown in red (positive polarity) and blue (negative polarity) patches. For reference, the coronal hole contours derived from SDO synoptic maps at 193 \AA\, are shown as grey patches. The HCS computed from the corresponding COCORIA simulation is shown in a thick red line.}
    \label{fig:connectivity_carrington}
\end{figure*} 

The full details of the magnetic connectivity along the COCORIA chain are explained in Paper~Ia, e.g., \cite{Kennis2024}. We recall here the main elements. 

The magnetic connectivity for a specific date (which we call a connectivity snapshot) is computed first by aligning the two MHD simulations. Indeed, as stated in Section~\ref{sec:mhd_models}, COCONUT and EUHFORIA have different coordinate frames, and thus need to be rotated to match one another. We also add an extra 10 degrees of longitudinal shift to account for the missing Coriolis and centrifugal forces in COCONUT. This value is the mean standard solid-body-correction described in \cite{Pomoell2018}, that was obtained empirically.

Once this is done, we can trace the magnetic field lines all the way from the Earth to the Sun: we use Earth coordinates in the HEEQ frame for the corresponding date to locate the Earth in the EUHFORIA simulation, then trace a first field line up to 0.1~AU. Then, we use this termination point as a new seed for COCONUT and trace the corresponding field line from 0.1~AU up to the solar surface. In Paper~Ia, we would then plot the corresponding termination point onto the input magnetogram to link it to photospheric magnetic structures. To compute the field lines, we use the PyVista Python package \citep{Sullivan2019}, which corresponds to a Runge-Kutta 2 integrator.

In Paper~Ia, we also introduced uncertainties linked to this computation. 
The first is temporal uncertainty: the solar wind takes several days to travel from the Sun to Earth, so connectivity is not instantaneous. To take this into account, we average connectivity points over the past few days before the event to capture the evolution of connectivity during the solar wind propagation phase. This temporal variation is due solely to a simple rotation of the simulation, so we assume that the magnetic field configuration did not vary over the studied period (which is not always the case at the maximum of activity).
The second type is spatial uncertainty: the magnetic field lines can be discontinuous or very divergent locally, so we include a possible deviation of a few degrees by computing several field lines for the connectivity, each starting from a seed point separated from Earth by an angle between 5 and 10 degrees in latitude and/or longitude. These angles were chosen because previous studies have shown that it is at these distances that we can start to see significant changes in structures in simulations \citep{Scolini2019, Pinto2021}. Ultimately, we use 9 points around the Earth's position and thus compute between 1 and 9 field lines, depending on their continuity. 
In the end, we take the mean position of the points and compute the standard deviation to obtain a magnetic connectivity zone rather than a magnetic connectivity point. If the difference in latitude exceeds 20 degrees, the zone is then split into two different sub-zones.

It is also possible to compute the velocity streamlines in complement to the magnetic field lines to get the stream connectivity. To be consistent with Paper~Ia, we will compute magnetic connectivity only in this study, but will examine stream connectivity in future work. 

\subsection{Connectivity for a full Carrington rotation}
\label{sec:connectivity_carrington}

In this study, instead of examining a connectivity snapshot on a specific date, we evaluate connectivity over a full Carrington rotation to assess the impact of the input map on the full inner heliosphere modelling. This means we have to change how we visualize our results. 

The new visualization is shown in Figure~\ref{fig:connectivity_carrington} for the full Carrington rotation number 2219 for the HMI synoptic map. The Earth's trajectory over the period is shown as a horizontal blue line. For each position, connectivity to a positive polarity open field area is shown with red lines, while connectivity to a negative polarity open field area is shown with blue lines, and connectivity to an equatorial open field area is shown with green lines. The open field areas of the corresponding COCORIA simulation are shown in red (positive polarity) and blue (negative polarity) patches. For reference, the coronal hole contours derived from SDO synoptic maps at 193 \AA\, are shown as grey patches. Here we make the distinction between coronal holes, which are the cold dark EUV features observed in the solar corona, and open magnetic field areas, which are easier to extract from numerical simulations. Though they often coincide and can be used as a proxy for one another, these two kinds of structures are not identical \citep{Perri2024}. The term \textit{coronal hole} will thus from now on refer only to contours from observations, while the term \textit{open field area} will refer only to contours from simulations. The HCS computed from the corresponding COCORIA simulation is shown in red. 
We recall that since we are focusing here on fast-solar-wind streams structure, the connecting source on the solar surface is most likely to be a coronal hole, which is why we do not include other solar features.
To facilitate interpretation of the figure, we have indicated the correspondence between longitude in degrees (bottom $x$-axis) and the date in the Carrington frame (top $x$-axis) when the Earth was at this longitude along its trajectory. 
Compared to Paper~Ia, each colored line now represents a connectivity snapshot at a given date. 

Regarding the uncertainties, we have adapted how we compute and display them. 
Since we plot two points per day, the temporal uncertainty is now reduced to 6 hours instead of four days. 
We continue to compute the spatial uncertainty at 5-degree and 10-degree separations, but we do not display it to avoid overwhelming the figure. The corresponding connectivity point in the open field area is an average of these spatial variations; we just do not display the associated standard deviation, although it is computed and accessible if needed for further analysis.  

\subsection{Expected connectivity from data}
\label{sec:ace}

With the selected period (Carrington rotation number 2219, which starts on June 29, 2019, and ends on July 26, 2019), we can look at in situ data from Earth to see what kind of connectivity we expect during this particular Carrington rotation. 

\begin{figure*}
    \centering  \includegraphics[width=\hsize]{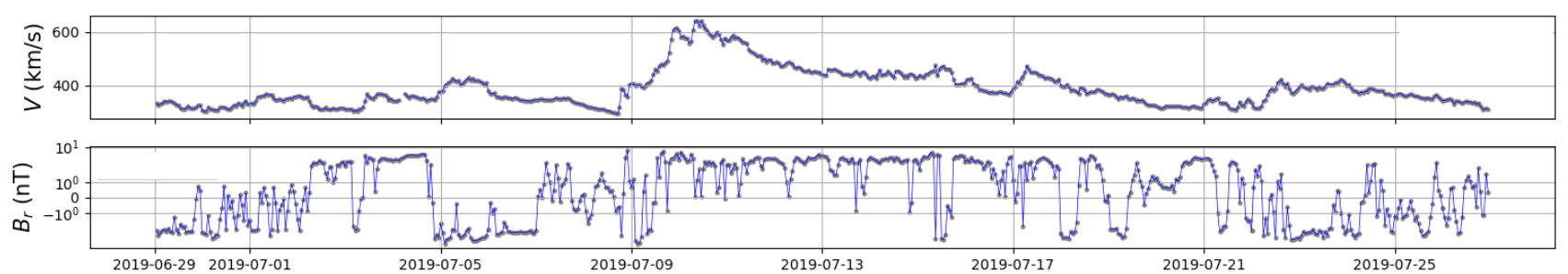}
    \caption{In-situ measures of the velocity of the solar wind in km/s (top panel) and the magnetic field in nanotesla (bottom panel) from the ACE spacecraft at L1. The results are presented for the studied period of Carrington rotation 2219 between 29/06/2019 and 26/07/2019.}
    \label{fig:insitu}
\end{figure*}

Figure~\ref{fig:insitu} shows the in-situ velocity (in km/s) and radial magnetic field (in nT) measured by the Advanced Composition Explorer (ACE) spacecraft at L1. 
We can see that over the period, there is a high-speed stream event between July 9 and July 13, associated with a positive equatorial coronal hole. If we put this information in perspective with Figure~\ref{fig:connectivity_carrington}, and especially with the grey patches corresponding to the observed coronal holes, this is likely a connection to the equatorial northern open field area visible between longitudes 250 and 300 (longitudes are given in degrees). 
For the rest of the period, connectivity fluctuates due to proximity to the HCS, but still shows periods of stability lasting several hours to two days. 

With the results of Papers Ia and Ib in mind, we know that although it is a period of minimum activity, it is quite a challenging configuration for connectivity. Indeed, the proximity of the HCS to Earth's trajectory can lead to small errors, resulting in an opposite-polarity result. 
We also know from Paper~Ib that the equatorial open field area connected to the high-speed stream is very challenging to reproduce, due to its proximity to the main active regions of the input maps. Thus, we do not expect many simulations to reproduce that result. Hence, instead of focusing on high-speed streams validation as we did in Paper~Ia, we will focus here instead on general HCS polarity validation, as described in \cite{Badman2022}. 

We note that choosing a longer validation period tends to improve the comparison with Carrington-frame maps. Indeed, since the synchronic-frame map is based on the date at the beginning of the Carrington rotation, and since they are not fully synchronic, some data may be outdated. However, since this is a minimum-activity configuration, we have checked that the magnetic field configuration and the coronal holes have not changed too much compared to the last Carrington rotation. Since the high-speed stream observed is indeed a recurring one (although it was slightly weaker during the previous Carrington rotation, with speeds reaching 500 km/s instead of 650 km/s), the comparison, even with maps that are not fully actualised, remains valid. It is also a good exercise in space weather forecasting to see whether we can use asynchronous data to predict connectivity for the next Carrington rotation. 
 
\section{Results}
\label{sec:results}

\begin{figure*}[!h]
\centering
  \includegraphics[width=\textwidth]{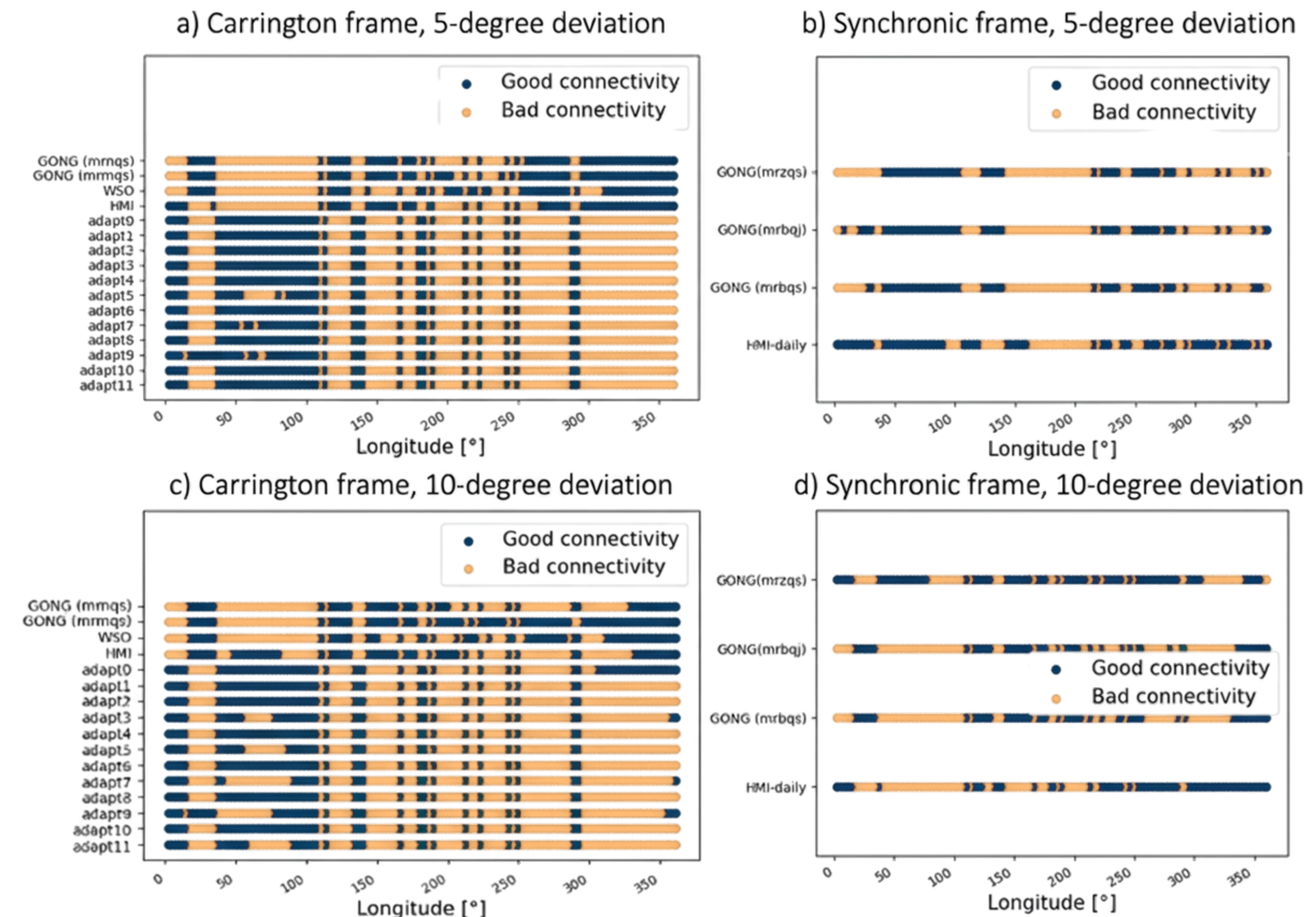}
  \caption{Visualization of the polarity agreement between simulations and observed data at Earth depending on the longitude (which corresponds to different dates), based on all input magnetic maps. The left panels are in the Carrington frame, while the right panels are in the synchronic frame. Dark blue indicates a good connectivity agreement; light orange indicates a bad connectivity agreement. A deviation of 5 was chosen for the top panels (lower spatial uncertainty) and a deviation of 10 for the bottom panels (higher spatial uncertainty).}
  \label{fig:connectivity_eval_total}
\end{figure*}

To present our results more comprehensively, we draw inspiration from the visualization described in \citet{Badman2022}. In their Fig. 3, they assessed the match between the simulated and measured polarity for the connectivity using a simple binary metric (agree/disagree) for each Carrington longitude of the studied period. This allowed them to derive a metric score as the percentage of agreeing polarity matches out of the total number of available matches.
We present a similar approach in Fig. \ref{fig:connectivity_eval_total}: for each longitude (which corresponds to a specific date in the corresponding synoptic map frame), if the simulated polarity at Earth agrees with the observed one, it is represented by a dark blue dot; mismatches are represented by a light orange dot. 
Here, the colors have nothing to do with the polarity sign (which was previously indicated by red for positive and blue for negative), just with the agreement. Also, it does not necessarily mean that we are connected to the correct open field area (which is very difficult to check), but at least that we are connected to an open field area of the correct polarity. 

In Figure~\ref{fig:connectivity_eval_total}, we have separated the cases with input maps set in the Carrington frame (GONG \textit{mrnqs} and \textit{mrmqs} products, WSO, HMI monthly, and ADAPT products) in the left column from the cases with input maps set in the synchronic frame (GONG \textit{mrzqs}, \textit{mrbqj} and \textit{mrbqs} products, and HMI daily products) in the right column. This means that their longitudes on the x-axis refer to different dates in the Carrington rotation. In the Carrington frame, longitudes refer to the beginning (June 29 2019) and end (July 26 2019) of CR2219 from the right side to the left side (from 360 to 0 degrees). In the synchronic frame, longitudes 0\degree to 120\degree include an update from the current day on the daily map (here, it was July 2nd 2019, based on the study in Paper~Ib). This means that in the synchronic frame, CR2219 begins at longitude 120\degree, then the dates increase from right to left, and when they reach longitude 0\degree, they circle back to longitude 360\degree. 

We have also separated connectivity estimates made with a 5-degree deviation around Earth to infer spatial uncertainties (top panel) with estimates made with a 10-degree deviation around Earth (bottom panel). If connectivity is consistent in this region, we do not expect any change as the deviation increases. If a change occurs, it indicates that the region around Earth has unstable connectivity, making it harder to provide a reliable estimate. 

We can start by analyzing qualitatively Figure~\ref{fig:connectivity_eval_total}. In panel a), we see that we get the connectivity right for this period for barely half of the dates only. This is not surprising, given the analysis of the data and Paper~Ia: minimum-activity cases are actually more challenging for MHD simulation chains because the Earth's position is close to the HCS, thereby amplifying any minor deviation of the HCS. This is also true in the data, as we can see frequent switching between positive and negative polarity between June 29 and July 3rd, then again between July 7 and July 10, and finally between July 22 and 26. 
We see that ADAPT products are highly consistent with one another, indicating minor differences across all 12 subproducts. However, they are not consistent with the other 4 maps from other providers. This result was also found in Paper~Ib: the magnetic flux accumulation at the poles gave different results with the same operational input parameters, in particular failing to reproduce the equatorial open field areas, which are key to the connectivity estimates. 
In panel b), we see that the synchronic frame maps tend to perform well with a 5-degree deviation. 
In panel c), we see that the Carrington-frame cases have quite robust connectivity estimates, as only the region between longitudes 50\degree and 100\degree and between 350\degree and 360\degree is affected by the increase in deviation around Earth, and only for certain maps (mainly ADAPT and HMI products). 
In panel d), we see that synchronic frame cases, on the other hand, are more sensitive to the deviation and thus have higher uncertainty, especially between longitudes 0\degree and 120\degree. This is expected, as the part of the frame filled with daily data, though more synchronic, is actually noisier and is more prone to geometric effects. 

There seem to be regions that are difficult to reconstruct across all the maps, corresponding to the dates stated above, where the HCS is very close to Earth's position. We will discuss this in more detail in the next section. 

For a more quantitative analysis, we convert this visual representation into a metric score to compare the results produced by each map. The results can be found in Table~\ref{tab:accuracy}.
To facilitate comparison, a ranking of the top five most favorable (in bold) and least favorable (in italics) results is presented. 
The top 5 comparisons for a 5-degree deviation include HMI daily (60.55\%), GONG mrmqs (57.79\%), GONG mrnqs (54.12\%), HMI monthly (50.45\%), and, very surprisingly, WSO (49.54\%), which is the lowest-resolution map. For the 10-degree deviation, the ranking changes slightly: GONG mrmqs becomes first with a slight score increase (63.30\%), GONG mrzqs becomes second with a big score increase (62.38\%), ADAPT0 ranks third with a massive score increase (53.21\%), HMI daily falls to fourth with a score decrease (51.37\%) and HMI monthly ranks fifth with the same score (50.45\%).  
This ranking is not globally surprising: from Paper~Ib, we know that HMI products yield the best results with the COCONUT routine input parameters, and, within the scope of the study, we know that the Carrington-frame maps are more likely to provide more consistent results over the entire period. The surprising results stem from WSO performing well on average: in Paper~Ib, it did not yield good results for coronal structures due to its low resolution it was ranked in the worst 3 scores for all the metrics we tested), but for connectivity, it seems sufficient for almost 50\% of the dates (which ranks it in the top maps for the time period studied). It is also interesting to find that one of the GONG-ADAPT maps yields a significantly better result with the 10-degree deviation, showing that even small differences can, in the end, produce different results.  
For the 5-degree deviation, the worst results are from the GONG-ADAPT maps, with scores failing to exceed 40\%. For the 10-degree deviation, we find the same trend, with the only exception of realisation 0.  

\begin{table}[!h]
    \caption{Accuracy percentage for the polarity estimate for 20 different input magnetic maps.}
    \centering
    \begin{tabular}{c|c|c|c|c}
    \hline 
    \hline 
Input & Deviation  & Ranking & Deviation & Ranking \\ 
map & of 5: [$\%$] & &  of 10: [$\%$] & \\
\hline 
HMI & 50.45 & 4 (+) & 50.45 & 5 (+)   \\
WSO & 49.54 & 5 (+) & 48.62 & 6    \\
mrbqs & 46.78 & 7 & 44.03 & 7  \\
mrzqs & 44.03 & 8 & 62.38 & 2 (+)   \\
mrnqs & 54.12 & 3 (+) & 39.44 & 9    \\
mrmqs & 57.79 & 2 (+) & 63.30 & 1 (+)    \\
mrbqj & 49.54 & 5 (+) & 41.28 & 8   \\
adapt0 & 36.69 & 10 & 53.21 & 3 (+)    \\
adapt1 & 36.69 & 10 & 36.69 & 10    \\
adapt2 & 36.69 & 10 & 36.69 & 10   \\
adapt3 & 36.69 & 10 & 33.02 & 16 (*)    \\
adapt4 & 36.69 & 10 & 36.69 & 10   \\
adapt5 & 29.35 & 20 (*) & 28.44 & 18 (*)   \\
adapt6 & 36.69 & 10 & 36.69 & 10   \\
adapt7 & 34.86 & 19 (*) & 24.77 & 20 (*)   \\
adapt8 & 36.69 & 10 & 36.69 & 10   \\
adapt9 & 38.53 & 9 & 33.02 & 16 (*)    \\
adapt10 & 36.69 & 10 & 36.69 & 10    \\
adapt11 & 36.69 & 10 & 28.44 & 18 (*)   \\
hmi-daily & 60.55 & 1 (+) & 51.37 & 4 (+)  \\
\hline 
\hline 
\end{tabular}
\tablefoot{The results were obtained for two different values for the deviation, namely 5 degrees and 10 degrees. For clarity, we present a ranking of scores from best (1) to worst (20). The top 5 scores are highlighted with a plus symbol (+), and the worst 5 scores are highlighted with a star symbol (*).}
\label{tab:accuracy}
\end{table}

The degree of accuracy varies with the selected deviation. In 10 of the 20 cases, the numerical results with a deviation of 5 degrees outperformed those with a deviation of 10 degrees when analysing the 20 different input magnetic maps. In seven out of the 20 cases, the results were identical. In contrast, in four cases, the results obtained with a 10-degree deviation were slightly better. This discrepancy can be attributed to the fact that when greater uncertainty is applied to Earth's location, it may result in an incorrect connection to the magnetic field line within the coronal region. Nevertheless, a rationale for selecting a deviation of 10 is that in the cases where greater accuracy is achieved, the level of accuracy is considerably higher. This indicates that the magnetic field lines in these simulations are subject to numerous perturbations, including magnetic reconnection that can go as far as to disconnect magnetic field lines from the Sun. It is therefore necessary to increase the deviation in order to avoid these perturbations and recover the large-scale trend. An accurate input magnetic map can be used to achieve higher levels of accuracy.

In conclusion, for the COCORIA chain used with non-optimised running parameters, it seems that the most important features of the input maps need to be consistent over the full Carrington rotation, even more so than high data resolution. 

\section{Discussion}
\label{sec:discussion}

\begin{figure*}
  \centering
  \includegraphics[width=\textwidth]{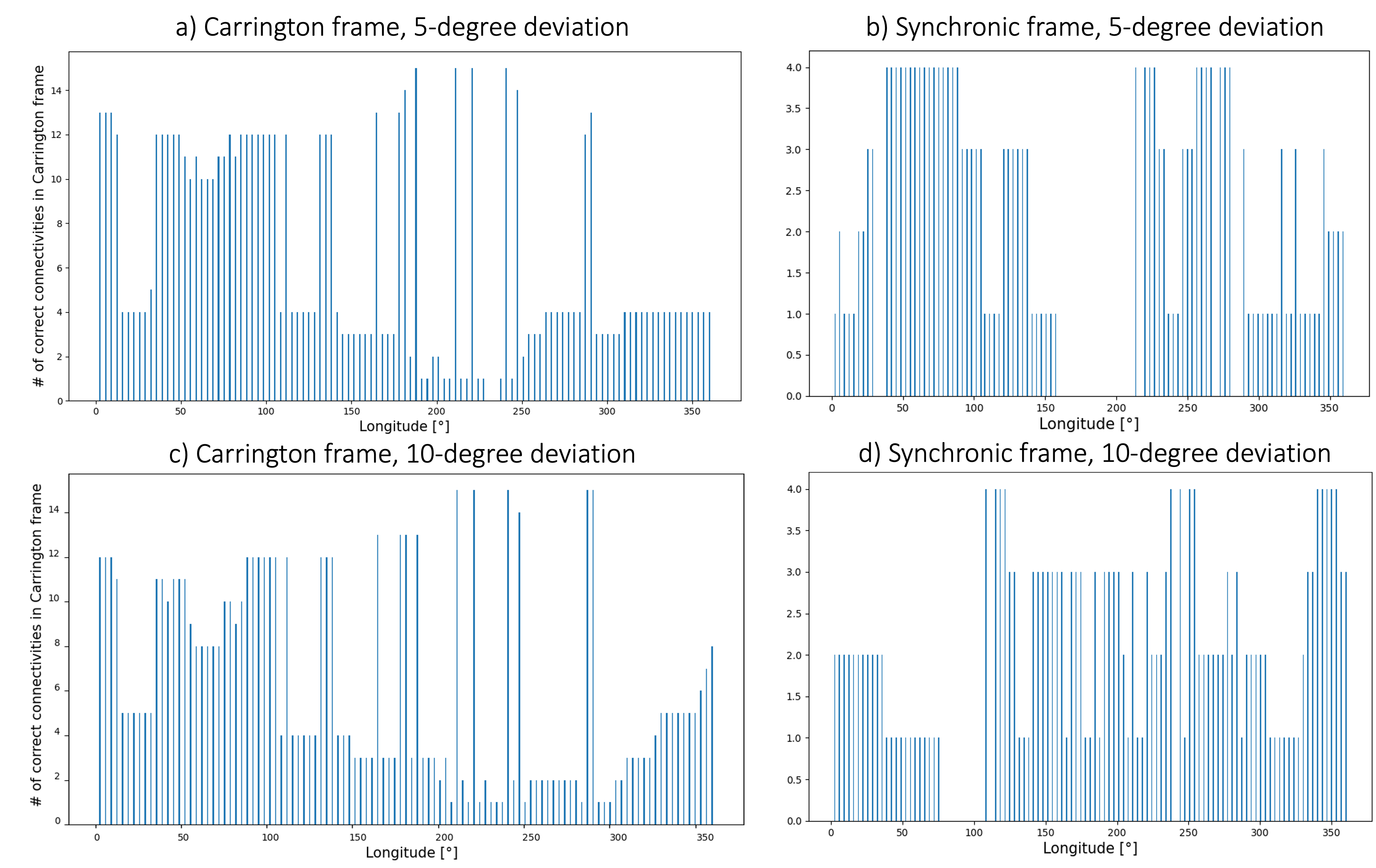}
\caption{Histograms of the numbers of correct polarity estimations depending on the longitude for the 20 input maps. The left panels present the outcomes of all the magnetic maps within the Carrington frame, while the right panels illustrate the outcomes of magnetic maps within the synchronic frame. The maximum value that can be reached is thus 16 for the Carrington frame and 4 for the synchronic frame. The results displayed in the top panels have been obtained using a 5-degree deviation around the Earth to account for spatial uncertainty, while the bottom panels have been obtained using a 10-degree deviation.}
  \label{fig:hist_long_tot}
\end{figure*}

To assess whether the poor connectivity estimates were due solely to the input map or to specific structures at particular longitudes that made the prediction more challenging, we have also plotted the distribution of good connectivity matches per longitude in Figure~\ref{fig:hist_long_tot} as histograms. In other words, for each longitude, we count how many times we have good agreement between the polarity measured in the in situ data and that computed in the simulation chain. We have separated Carrington (left) from synchronic (right) frames, which means that the maximum number of counts that can be reached for Carrington-frame plots is 16 vs. 4 for the synchronic-frame ones. We have separated the results for spatial uncertainty induced by a 5-degree (top row) vs 10-degree (bottom row) deviation from Earth's position. 
For the Carrington-frame maps, the most challenging longitudes correspond to 200 to 300 degrees (less than 4 matches overall), while the most consistent longitudes are between 0 and 100 degrees (on average, more than 8 matches, which is half the sample of maps). These results are consistent whether you use a 5-degree or 10-degree deviation. 
For the synchronic-frame maps however, we have already seen that they are more sensitive to the uncertainty deviation, so the results are going to be different for the 5-degree vs 10-degree deviation: in panel b), there is a clear gap of good connectivity estimates between 150 and 210 degrees (with 0 match most of the time), while in panel d), the longitudes 0\degree to 100\degree are the most affected. In the Carrington frame, those correspond, respectively, to regions between longitudes 30 and 90 degrees and 260 and 360 degrees. This means that the challenging longitudes do not overlap for the 5-degree deviation, but they do for the 10-degree deviation. 

The region between longitudes 200\degree and 360\degree in Carrington frame is expected to be more challenging: in Paper~Ib, we see a massive active region during this Carrington rotation that is causing the opening of equatorial coronal holes. The other regions are more surprising, but actually match periods in the in situ data where the Earth is very close to the HCS position. 
From the in situ data, we expect stable connectivity to the fast solar wind stream from longitudes 130\degree to 230\degree. These longitudes did not consistently yield good scores across all maps, indicating that this connectivity was not sustained over the full period and was strongly affected by the choice of the input map. 

It is important to note that the occurrence of a significant polarity switch does not necessarily indicate an incorrect magnetic connection. In fact, there are instances in which the polarity of the in situ data changes substantially but does not affect the outcome. Nevertheless, it is essential to consider the possibility that these regions may prove particularly challenging to model, even when employing MHD modelling. 

\section{Conclusion}
\label{sec:conclusion}

This project was a follow-up of \cite{Kennis2024} (Paper~Ia), which studied magnetic connectivity from the Sun to the Earth with MHD models. We decided to use the same approach as in \cite{Perri2023} (Paper~Ib), in which we assessed the impact of the input magnetic map on the estimation of magnetic connectivity. Therefore, we used the same 20 distinct input magnetic maps for Carrington Rotation number 2215 (on July 2nd 2019, more exactly for the daily-updated products). We had already assessed the quality of the coronal description produced by the COCONUT model in Paper~Ib, and this time we coupled these solutions with EUHFORIA to propagate the information all the way to Earth. A new approach from Papier Ia was to assess connectivity over a longer time period (the full Carrington rotation instead of a few days), which had not been done previously with MHD models. The accuracy of the results was assessed by comparing the simulated data with the in situ data polarity, following an approach similar to \cite{Badman2022}. This approach allows us to draw conclusions about the location of the magnetic connectivity point relative to the HCS.  

From the results and discussion section, we can summarize our main results as follows: 
\begin{itemize}
    \item Unsurprisingly, since the choice of the input map has strong consequences on the coronal model's ability to reproduce key coronal structures, it has a strong impact on the final connectivity estimates. We were able to assess this effect quantitatively in a "simple" case at a minimum level of activity: the change in the input map alone can impact the connectivity estimation during this period, going from 28\% global agreement to 63\% in the most extreme case. It is thus essential for all connectivity models to try all available maps to determine which one consistently yields the best results, or to adapt their parameters to each map separately.
    \item Following Paper~Ib, as the HMI maps were providing the best comparison with coronal observations, they are also very good products to estimate the magnetic connectivity using the COCORIA chain (with scores consistently above 50\% of agreement over the full time period, placing them among the top 5 maps, no matter the spatial uncertainty). Similarly, because the ADAPT maps do not yield satisfactory results for the coronal model under the operational parameters of the COCORIA chain, they also yield low average connectivity scores (mostly below 37\%). But the correspondence between coronal and connectivity performances is not always one-to-one: the WSO maps provided bad estimates for the coronal properties in Paper~Ib, but provide actually very good results for the connectivity in this paper. This means that, although the connectivity score can, on average, be linked to the prediction of coronal properties, it is not always so direct and should be consistently checked, especially for low-resolution magnetic maps. 
    \item For the COCORIA chain, we find that the most confident way to get good connectivity estimates over a long period is to have the most consistent and calibrated data over the full Carrington rotation. Thus, monthly products are more reliable than daily products for this kind of study. However, it may be possible in the near future to go beyond this limitation by using real synchronic maps based on Solar Orbiter far-side data \citep{Perri2024}.
    \item Following the conclusions from Papers Ia and Ib, we also find that minimum-activity cases are actually challenging for MHD models and connectivity in general, because these are periods where the Earth is very close to the HCS, and thus small uncertainties can lead to wrong polarity estimations. Thus, especially at minimum activity, it is crucial to assess the impact of the input magnetic map, which is likely to be the main factor of uncertainties and errors. 
    \item This study shows that, for operational predictions, it would be optimal to adopt time-dependent boundary conditions that integrate more monthly information into the daily products, which will be the focus of a follow-up study based on the work of \cite{Baratashvili2025}, similar to what has been done in \cite{Lionello2026}. Continued numerical improvements may also provide better results (like including more accurate integrators into COCONUT and EUHFORIA), especially in regions where the solution is rapidly changing. Finally, experimenting with one MHD model instead of two could be interesting to minimize the propagation of error, and is now possible using only the COCONUT model \citep{Wang2026}.
\end{itemize} 

\begin{acknowledgements}
    The authors thank Rui Pinto, Predictive Science Inc. and the ISWAT S2-05 team for valuable discussions. 
    This work received funding from the French National Research Agency under reference ANR-23-ASZC-0004 (ANR Astrid WindTRUST).
    SK and SP acknowledge support from the projects C16/24/010 (C1 project Internal Funds KU Leuven), G0B5823N and G002523N (WEAVE) (FWO-Vlaanderen), and 4000145223 (SIDC Data Exploitation (SIDEX2). 
    SP is funded by the European Union (ERC-AdG agreement No 101141362, Open SESAME). Views and opinions expressed are, however, those of the author(s) only and do not necessarily reflect those of the European Union or the European Research Council. Neither the European Union nor the granting authority can be held responsible for them.
    We used the VSC–Flemish Supercomputer Centre infrastructure for the computations, funded by the Hercules Foundation and the Flemish Government, Department EWI. 
    We thank the ACE SWEPAM instrument team and the ACE Science Center for providing the ACE data.
    Wilcox Solar Observatory data used in this study were obtained via the website \url{http://wso.stanford.edu} courtesy of J.T.\ Hoeksema.
    The Wilcox Solar Observatory is currently supported by NASA.
    Data were acquired by GONG instruments operated by NISP/NSO/AURA/NSF with contributions from NOAA.
    HMI data are courtesy of the Joint Science Operations Center (JSOC) Science Data Processing team at Stanford University.
    This work utilises data produced collaboratively between AFRL/ADAPT and NSO/NISP.
    Data used in this study was obtained from the following websites: \\
    WSO: \url{http://wso.stanford.edu/synopticl.html} \\
    GONG: \url{https://gong2.nso.edu/archive/patch.pl?menutype=z} \\
    HMI: \url{http://jsoc.stanford.edu/HMI/LOS_Synoptic_charts.html} \\
    GONG-ADAPT: \url{https://gong.nso.edu/adapt/maps/}
\end{acknowledgements}

\bibliographystyle{aa} 
\bibliography{aa62134-26}
\end{document}